\documentclass[preprint,showpacs,preprintnumbers,amsmath,amssymb]{revtex4}
\usepackage{graphics}
\usepackage{dcolumn}
\usepackage{bm}

\begin{document}

\title{Angular momentum of non-paraxial light beam: Dependence of orbital angular momentum
on polarization}

\author{Chun-Fang Li\footnote{Email address: cfli@shu.edu.cn}}

\affiliation{Department of Physics, Shanghai University, 99 Shangda Road, Baoshan District, 200444
Shanghai, P. R. China}

\date{\today}

\begin{abstract}

It is shown that the momentum density of free electromagnetic field splits into two
parts. One has no contribution to the net momentum due to the transversality condition.
The other yields all the momentum. The angular momentum that is associated with the
former part is spin, and the angular momentum that is associated with the latter part is
orbital angular momentum. Expressions for the spin and orbital angular momentum are given
in terms of the electric vector in reciprocal space. The spin and orbital angular
momentum defined this way are used to investigate the angular momentum of nonparaxial
beams that are described in a recently published paper [Phys. Rev. A 78, 063831 (2008)].
It is found that the orbital angular momentum depends, apart from an $l$-dependent term,
on two global quantities, the polarization represented by a generalized Jones vector and
a new characteristic represented by a unit vector $\mathbf{I}$, though the spin depends
only on the polarization. The polarization dependence of orbital angular momentum through
the impact of $\mathbf{I}$ is obtained and discussed. Some applications of the result
obtained here are also made. The fact that the spin originates from the momentum density
that has no contribution to the net momentum is used to show that there does not exist
the paradox on the spin of circularly polarized plane wave. The polarization dependence
of both spin and orbital angular momentum is shown to be the origin of conversion from
the spin of a paraxial Laguerre-Gaussian beam into the orbital angular momentum of the
focused beam through a high numerical aperture.

\end{abstract}

\maketitle


\section{Introduction}

The orbital angular momentum (AM) of light did not draw much attention \cite{Santamato,
Abbate} until 1992 when Allen and his co-researchers \cite{Allen-BSW} showed that a beam
of Laguerre-Gaussian mode can carry both spin and orbital AM. They found that the spin is
carried by the polarization $\sigma$ and the orbital AM is carried by the helical wave
front represented by a phase factor $\exp(il\phi)$, where $l$ is an integer. Since then
great progress has been made \cite{Franke-Arnold} in experiments. The orbital AM has been
measured \cite{Leach, Berkhout-B}. The transfer of spin and orbital AM to microscopic
particles \cite{Simpson-DAP, Friese, He-FHR, Paterson} and to liquid crystals
\cite{Piccirillo, Piccirillo1} has been observed.

Recently, experimentalists \cite{O'Neil, Garces} showed that the spin and orbital AM of a
non-paraxial beam play distinct roles in the interaction with microscopic birefringent
particles trapped off the beam axis in optical tweezers. It was observed \cite{Garces}
that the spin of light makes the particle rotate around its own axis and the orbital AM
makes the particle rotate around the beam's axis. Furthermore, partial spin of a paraxial
beam was observed \cite{Zhao} to be converted into orbital AM of a non-paraxial beam by a
high numerical aperture. Those experimental results demonstrate that the spin and orbital
AM of a non-paraxial beam are different in nature on one hand and are connected somehow
to each other on the other. But up till now, there is no satisfactory theory to elucidate
the difference and relation. The distinction that the spin is carried by the polarization
and the orbital AM is carried by the helical wave front was drawn basically from the
knowledge of a type of paraxial beams \cite{Allen-BSW, van1, Berry}. It is not valid for
non-paraxial beams \cite{Barnett-A, Allen-MM, Barnett}. With a specific non-paraxial
beam, Barnett and Allen \cite{Barnett-A} found that ``the seemingly natural separation of
the angular momentum...is no longer possible''. The purpose of this paper is to advance a
theory to explain the difference and relation between the spin and orbital AM of
nonparaxial beams.

To this end, we should first know how to represent a nonparaxial beam that as a whole is
in a definite state of polarization. As mentioned before, Barnett and Allen
\cite{Barnett-A} once put forward a nonparaxial solution. But that solution was shown
\cite{Li} to fail to meet the demand. Fortunately, a theoretical representation that
meets the demand was recently developed \cite{Li1}. The beam in this representation
exhibits as a whole a definite polarization in the sense that all the plane waves that
constitute the beam are described by the same normalized Jones vector. In other words,
the normalized Jones vector in this representation is a global characteristic that plays
the role of describing the polarization of the beam. This Jones vector will be referred
to as the generalized Jones vector. Apart from the global generalized Jones vector, a
non-paraxial beam in this representation shows another global characteristic denoted by a
unit vector. The global unit vector was applied \cite{Li3} to explain the spin Hall
effect of light \cite{Hosten-K}. In this paper, I will make use of this representation to
show how the orbital AM depends on the polarization through the impact of the global unit
vector.

Secondly, we should also know how to define the spin and orbital AM of an electromagnetic
field in free space. The total AM $\mathbf{J}(\mathbf{x}_0)$ of a free electromagnetic
field with respect to the point $\mathbf{x}_0$ is defined as \cite{Mandel}
\begin{equation} \label{TAM x0}
\mathbf{J}(\mathbf{x}_0)= \int \mathbf{j} d^3 x= \mathbf{J}(0) -\mathbf{x}_0 \times \int
\mathbf{p} d^3 x,
\end{equation}
where $\mathbf{j}= (\mathbf{x}-\mathbf{x}_0) \times \mathbf{p}$ is the AM density with
respect to the same reference point, $\mathbf{p}=\varepsilon_0 \mu_0 \mathcal{E} \times
\mathcal{H}$ is the momentum density defined in terms of the electric vector
$\mathcal{E}$ and the magnetic vector $\mathcal{H}$, and
\begin{equation} \label{TAM origin}
\mathbf{J}(0)= \int \mathbf{x} \times \mathbf{p} d^3 x
\end{equation}
is the AM with respect to the origin. The separation of total AM into spin and orbital AM
was discussed before \cite{Humblet, Cohen, van, Mandel} by performing the integration in
Eq. (\ref{TAM x0}) by parts and neglecting a surface integral at infinity. In this paper,
I will put forward a rigorous approach to the separation of total AM into spin and
orbital parts by examining the property of momentum density. This approach allows us to
apply the obtained result to plane waves.

The paper is arranged as follows. In Section \ref{AM separation}, it is found from the
transversality condition that the momentum density of an electromagnetic field in free
space splits into two parts. One part does not have any contribution to the net momentum;
the other part produces all the momentum. The AM that originates from the former part
does not depend on the choice of the reference point and is the spin. The AM that
originates from the latter part is in general dependent on the choice of the reference
point and is the orbital AM. In Section \ref{AM-propagating beam}, the integral
expressions for the spin and orbital AM obtained in Section \ref{AM separation} are used
to investigate the AM properties of nonparaxial beams described by the aforementioned
representation. Since the light beam is assumed to be monochromatic, both the integrals
of spin and orbital AM are infinite. In order to deal with the infinity, the technique of
$\delta$-function normalization is used. As expected, the spin AM is found to be
dependent on the polarization. But what is surprising is that the orbital AM is also
dependent on the polarization. It is shown how the orbital AM depends on the polarization
through the impact of the global unit vector. Two different problems are discussed in
Section \ref{Applications} by making use of the obtained results. Section
\ref{conclusions} concludes the paper with further remarks.

\section{Separation of the total AM into spin and orbital AM} \label{AM separation}

Consider an arbitrary electromagnetic field in free space. Its electric vector in real
space can be expressed as an integral over the plane-wave spectrum,
\begin{equation}\label{FE-E}
\mathcal{E}(\mathbf{x},t)=\frac{1}{2} \left\{\frac{1}{(2 \pi)^{3/2}} \int \mathbf{E}(\mathbf{k})
\exp[i(\mathbf{k} \cdot \mathbf{x}- \omega t)] d^3 k+ c. c. \right\},
\end{equation}
where $\mathbf{k}$ is the wave vector and $\mathbf{E}(\mathbf{k})$ is the electric vector
in reciprocal space. The magnetic vector of the beam is derived from Eq. (\ref{FE-E}) and
Maxwell's equations to be
\begin{equation} \label{FE-H}
\mathcal{H}(\mathbf{x},t)=\frac{1}{2} \left\{ \frac{1}{(2 \pi)^{3/2}} \int
\frac{\mathbf{k} \times \mathbf{E}}{\mu_0 \omega} \exp[i(\mathbf{k} \cdot \mathbf{x}-
\omega t)] d^3 k+ c.c. \right\}.
\end{equation}
Integral expression (\ref{FE-E}) or (\ref{FE-H}) leads to the following transformations
\cite{Mandel},
\begin{equation}\label{Hermitian}
\omega(-\mathbf{k})= -\omega(\mathbf{k}), \hspace{10pt} \mathbf{E}(-\mathbf{k})=
\mathbf{E}^* (\mathbf{k}).
\end{equation}
With the help of Eqs. (\ref{FE-E}) and (\ref{FE-H}) and vector algebra $\mathbf{a} \times
(\mathbf{b} \times \mathbf{c})= (\mathbf{a} \cdot \mathbf{c}) \mathbf{b}- (\mathbf{a}
\cdot \mathbf{b}) \mathbf{c}$, the momentum density splits into two parts,
\begin{equation}
\mathbf{p}=\varepsilon_0 \mu_0 \mathcal{E} \times \mathcal{H}= \mathbf{p}_1+
\mathbf{p}_2,
\end{equation}
where
\begin{eqnarray} \label{part 1 of p}
\mathbf{p}_1 & = & \frac{\varepsilon_0}{4(2 \pi)^3}
                   \int \frac{\mathbf{E}' \cdot \mathbf{E}}
                   {\omega} \mathbf{k} e^{i(\mathbf{k}'+ \mathbf{k}) \cdot \mathbf{x}}
                   e^{-i(\omega'+ \omega)t} d^3 k' d^3 k \nonumber \\
             &   & +\frac{\varepsilon_0}{4(2 \pi)^3}
                   \int \frac{\mathbf{E}' \cdot \mathbf{E}^*}
                   {\omega} \mathbf{k} e^{i(\mathbf{k}'- \mathbf{k}) \cdot \mathbf{x}}
                   e^{-i(\omega'- \omega)t} d^3 k' d^3 k+ c.c. , \\
\label{part 2 of p}
\mathbf{p}_2 & = & -\frac{\varepsilon_0}{4 (2 \pi)^3}
                    \int \frac{\mathbf{E}' \cdot \mathbf{k}}
                    {\omega} \mathbf{E} e^{i(\mathbf{k}'+ \mathbf{k}) \cdot \mathbf{x}}
                    e^{-i(\omega'+ \omega)t} d^3 k' d^3 k \nonumber \\
             &   & -\frac{\varepsilon_0}{4 (2 \pi)^3}
                    \int \frac{\mathbf{E}' \cdot \mathbf{k}}
                    {\omega} \mathbf{E}^* e^{i(\mathbf{k}'- \mathbf{k}) \cdot \mathbf{x}}
                    e^{-i(\omega'- \omega)t} d^3 k' d^3 k+c.c. ,
\end{eqnarray}
$\mathbf{E} \equiv \mathbf{E}(\mathbf{k})$, $\mathbf{E}' \equiv \mathbf{E}(\mathbf{k}')$,
$\omega \equiv \omega(\mathbf{k})$, and $\omega' \equiv \omega(\mathbf{k}')$. Based on
the transversality condition $\mathbf{k} \cdot \mathbf{E}=0$, it is readily proven by use
of transformations (\ref{Hermitian}) that $\mathbf{p}_2$ does not have any contribution
to the net momentum,
\begin{equation} \label{null momentum part}
\mathbf{P}_2= \int \mathbf{p}_2 d^3 x=0.
\end{equation}
This tells us a fact that all the momentum $\mathbf{P}$ comes only from $\mathbf{p}_1$,
\begin{equation} \label{total momentum}
\mathbf{P}= \mathbf{P}_1= \int \mathbf{p}_1 d^3 x= \varepsilon_0 \int \frac{\mathbf{E}^*
\cdot \mathbf{E}}{\omega} \mathbf{k} d^3 k,
\end{equation}
which is independent of time.

Accordingly, the total AM also splits into two parts,
$$
\mathbf{J}(\mathbf{x}_0)= \int (\mathbf{x}- \mathbf{x}_0) \times \mathbf{p} d^3 x=
\mathbf{S} (\mathbf{x}_0)+ \mathbf{L} (\mathbf{x}_0).
$$
Because of property (\ref{null momentum part}), the first part $\mathbf{S}$ that
originates from momentum density $\mathbf{p}_2$ is independent of the choice of the
reference point,
\begin{equation} \label{total SAM}
\mathbf{S}(\mathbf{x}_0)= \mathbf{S}(0)= \int \mathbf{x} \times \mathbf{p}_2 d^3 x.
\end{equation}
In other words, the fact that $\mathbf{S}$ is independent of the choice of the reference
point roots in an intrinsic property of the electromagnetic field, the transversality
condition. It is thus reasonable to regard this intrinsic AM as the spin. The second part
that originates from momentum density $\mathbf{p}_1$ is in general dependent on the
choice of the reference point,
\begin{equation} \label{OAM wrt x0}
\mathbf{L}(\mathbf{x}_0)= \mathbf{L}(0)- \mathbf{x}_0 \times \mathbf{P}_1,
\end{equation}
where
\begin{equation} \label{total OAM}
\mathbf{L}(0)= \int \mathbf{x} \times \mathbf{p}_1 d^3 x.
\end{equation}
It is plausible to regard this part as the orbital AM. Substituting Eq. (\ref{part 2 of
p}) into Eq. (\ref{total SAM}), one obtains by straightforward calculations
\begin{equation} \label{S}
\mathbf{S}= \int \frac{\varepsilon_0}{i \omega} \mathbf{E}^* \times \mathbf{E} d^3 k.
\end{equation}
The momentum density $\mathbf{p}_2$ leads to the spin AM, though it does not produce any
momentum. Such an astonishing fact means that there is no paradox on the spin AM of
circularly polarized plane waves. This will be discussed in Section \ref{Applications}.
Substituting Eq. (\ref{part 1 of p}) into Eq. (\ref{total OAM}), one has
\begin{equation} \label{L}
\mathbf{L}(0)= \int \frac{\varepsilon_0}{i \omega} \mathbf{E}^{\dag} (\mathbf{k} \times
\nabla_{\mathbf{k}}) \mathbf{E} d^3 k,
\end{equation}
where $\nabla_{\mathbf{k}}$ is the gradient operator with respect to $\mathbf{k}$, and
the superscript $\dag$ stands for the conjugate transpose \cite{comment}. For the
readers' convenience, the details to calculate Eqs. (\ref{S}) and (\ref{L}) are
summarized in Appendix. It is very interesting to note that the spin (\ref{S}) and
orbital AM (\ref{L}) obtained this way look very like their quantum-mechanical
counterparts \cite{Mandel}.

At last, let us give here for later convenience the total energy of the beam in terms of
the plane-wave spectrum,
\begin{equation} \label{W}
W= \int (\frac{\varepsilon_0}{2} \mathcal{E}^{\dag} \mathcal{E}+ \frac{\mu_0}{2}
\mathcal{H}^{\dag} \mathcal{H}) d^3 x
 = \int \varepsilon_0 \mathbf{E}^{\dag} \mathbf{E} d^3 k.
\end{equation}

\section{\label{AM-propagating beam} AM properties of non-paraxial beams}

The AM of a propagating beam in the $z$-direction is commonly considered in the
literature \cite{Allen-BSW, Berry, Barnett-A, van1, Allen-MM} to be equivalent to the
line density, that is to say, to the AM per unit length in the $z$-direction. In order to
avoid any possible ambiguity that may arise from the AM density \cite{Allen-P, Monteiro},
I do not use this notion here. In fact, we have given in Eqs. (\ref{S}) and (\ref{L}) the
expressions for the spin and orbital AM themselves with respect to the origin. In this
section, we will use those expressions to investigate the AM properties of nonparaxial
beams. To do this, let us now convert the representation form of nonparaxial beams that
was advanced in Ref. \cite{Li1} into a form that is suitable for present purpose.

\subsection{Description of non-paraxial beams: introduction to a new global unit vector}

The electric vector $\mathcal{E}$ of a nonparaxial beam in real space is given by Eq.
(\ref{FE-E}). The electric vector $\mathbf{E}$ in reciprocal space is factorized into
three factors \cite{Li1},
\begin{equation} \label{factorization}
\mathbf{E}= m \tilde{\alpha} f,
\end{equation}
where
\begin{equation} \label{MM}
m=(\begin{array}{lr} \mathbf{u} & \mathbf{v} \end{array})
\end{equation}
is the mapping matrix, $\tilde{\alpha}= \left( \begin{array}{c} \alpha_1 \\
\alpha_2 \end{array} \right)$ is the generalized Jones vector that is assumed to be
independent of the wave vector and to satisfy the normalization condition
$\tilde{\alpha}^{\dag} \tilde{\alpha}=1$, and $f$ is the electric scalar in reciprocal
space. The unit column vectors $\mathbf{u}$ and $\mathbf{v}$ of $m$ represent the two
mutually orthogonal states of linear polarization and are defined in terms of the local
wave vector $\mathbf{k}$ and a global unit vector $\mathbf{I}$ as follows,
\begin{equation} \label{u and v}
\mathbf{u}= \mathbf{v} \times \frac{\mathbf{k}}{k}, \hspace{5pt} \mathbf{v}=
\frac{\mathbf{k} \times \mathbf{I}}{|\mathbf{k} \times \mathbf{I}|},
\end{equation}
which lead to an important normalization property of the mapping matrix,
\begin{equation} \label{normalization}
m^T m=1,
\end{equation}
where the superscript $T$ denotes the transpose. Unit vector $\mathbf{I}$ can be
specified by its polar angle $\Theta$ and azimuthal angle $\Phi$. For the sake of
simplicity, let us assume $\mathbf{I}$ to lie in the plane $zox$, that is to say
$\Phi=0$. In this case, we have
$$
\mathbf{I} (\Theta)= \mathbf{e}_x \sin \Theta+
\mathbf{e}_z \cos \Theta
$$
and the mapping matrix
\begin{equation}
m= \frac{1}{k |\mathbf{k} \times \mathbf{I}|}
   \left(
         \begin{array}{cc}
                 (k_y^2+k_z^2) \sin \Theta- k_z k_x \cos \Theta &  k k_y \cos \Theta \\
                -k_y (k_z \cos\Theta+ k_x \sin\Theta)           &  k (k_z \sin\Theta- k_x \cos\Theta) \\
                 (k_x^2+k_y^2) \cos \Theta- k_z k_x \sin \Theta & -k k_y \sin \Theta
         \end{array}
   \right),
\end{equation}
where $|\mathbf{k} \times \mathbf{I}|= [k^2- (k_x \sin \Theta+ k_z \cos
\Theta)^2]^{1/2}$. Due to the symmetry relation $\mathbf{I} (\Theta+ \pi)= -\mathbf{I}
(\Theta)$, it is postulated throughout this paper that
\begin{equation} \label{postulation}
|\Theta| \leq \frac{\pi}{2}.
\end{equation}

A monochromatic beam has a definite wavenumber. It is convenient to use spherical polar
coordinates to express the electric scalar as
$$
f= \frac{\delta(k-k')}{k^2} \bar{f}(\vartheta,\varphi),
$$
where $0 \leq \vartheta \leq \pi$ and $\hspace{5pt} 0 \leq \varphi \leq 2\pi$. Since
$\bar{f} (\vartheta, \varphi)$ is a periodic function of $\varphi$ with period $2 \pi$, a
physically allowed function has the following Fourier expansion:
$$
\bar{f}(\vartheta,\varphi)= \sum_{l=-\infty}^{\infty} f_l (\vartheta) \exp(il \varphi).
$$
In this paper, we consider only one term of the expansion and rewrite the electric scalar
as follows,
\begin{equation} \label{f in s polar}
f= \frac{\delta(k-k')}{k^2} f_l (\vartheta) \exp (il \varphi),
\end{equation}
where the angular-spectrum function $f_l (\vartheta)$ is assumed to be square integrable.
In order to use the technique of $\delta$-normalization, the complex conjugate of $f$ is
replaced with
\begin{equation} \label{f* in s polar}
f^*= \frac{\delta(k-k'')}{k^2} f^*_l (\vartheta) \exp (-il \varphi).
\end{equation}
For a beam that propagates in the $z$-direction, its angular-spectrum function satisfies
\begin{equation} \label{fl propagation}
f_l (\vartheta)=0 \hspace{5pt} \text{for} \hspace{5pt} \frac{\pi}{2} \leq \vartheta \leq
\pi.
\end{equation}
Furthermore, if the beam is well-collimated and thus can be paraxially approximated,
$|f_l (\vartheta)|$ is sharply peaked at $\vartheta=0$. The half width $\Delta \vartheta$
of $|f_l (\vartheta)|$ is the divergence angle of the beam.

So obtained $\mathbf{E}$ guarantees that the field vectors $\mathcal{E}$ and
$\mathcal{H}$ in Eqs. (\ref{FE-E}) and (\ref{FE-H}) satisfy Maxwell's equations. Now that
unit real vectors $\mathbf{u}$ and $\mathbf{v}$ are orthogonal to each other, the
$\tilde{\alpha}$ that is independent of the wave vector acts as a global characteristic
to describe the inner degree of freedom of the beam, the state of polarization. We thus
have two independent global quantities, $\mathbf{I}$ and $\tilde{\alpha}$, to describe a
beam. It should be pointed out that a physically allowed beam may be a linear
superposition of a series of so described beam. They each have their own $\mathbf{I}$ and
$\tilde{\alpha}$. The beam that we will consider in this paper is assumed to have
definite $\mathbf{I}$ as well as $\tilde{\alpha}$. In the following, we will pay much
attention to the effect of these two global characteristics on the orbital AM. Only the
AM with respect to the origin will be considered.

\subsection{Orbital AM is dependent on $\mathbf{I}$ as well as $\sigma$}

The longitudinal component of orbital AM with respect to the origin can be turned from
Eq. (\ref{L}) into
\begin{equation} \label{Lz in pw spectrum}
L_z= \int \frac{\varepsilon_0}{\omega} \mathbf{E}^{\dag} (-i \frac{\partial}{\partial
\varphi}) \mathbf{E} k^2 \sin \vartheta dk d\vartheta d\varphi
\end{equation}
in spherical polar coordinates. Hereafter the symbol for the origin will be omitted for
the sake of simplicity. By making use of Eq. (\ref{factorization}), one has
\begin{equation} \label{product}
\mathbf{E}^{\dag} (-i \frac{\partial \mathbf{E}}{\partial \varphi})
  =\tilde{\alpha}^{\dag} m^{\dag} (-i \frac{\partial m}{\partial \varphi})
  \tilde{\alpha} f^* f+ f^* (-i \frac{\partial f}{\partial \varphi}).
\end{equation}
When property (\ref{normalization}) is taken into account, straightforward calculations
yield
\begin{eqnarray} \label{derivative}
m^{\dag} (-i \frac{\partial m}{\partial \varphi}) &
  = & -\hat{\sigma}_3 \cos \vartheta
  +\frac{\hat{\sigma}_3}{2} \frac{\cos \vartheta- \cos \Theta}
  {1- \cos \Theta \cos \vartheta- \sin \Theta \sin \vartheta \cos \varphi} \nonumber \\
 & &
  +\frac{\hat{\sigma}_3}{2} \frac{\cos \vartheta+ \cos \Theta}
  {1+ \cos \Theta \cos \vartheta+ \sin \Theta \sin \vartheta \cos \varphi},
\end{eqnarray}
where $\hat{\sigma}_3= \left( \begin{array}{cc} 0 & -i \\ i & 0 \end{array} \right)$ is
the Pauli matrix. Substituting Eq. (\ref{derivative}) into Eq. (\ref{product}) and
noticing Eq. (\ref{f in s polar}), one obtains
\begin{eqnarray}
\mathbf{E}^{\dag} (-i \frac{\partial \mathbf{E}}{\partial \varphi}) &
  = & (l- \sigma \cos \vartheta) f^* f
  +\frac{\sigma}{2} \frac{(\cos \vartheta- \cos \Theta) f^* f}
  {1- \cos \Theta \cos \vartheta- \sin \Theta \sin \vartheta \cos \varphi} \nonumber \\
 & &
  +\frac{\sigma}{2} \frac{(\cos \vartheta+ \cos \Theta) f^* f}
  {1+ \cos \Theta \cos \vartheta+ \sin \Theta \sin \vartheta \cos \varphi}. \nonumber
\end{eqnarray}
Substituting it into Eq. (\ref{Lz in pw spectrum}) and considering Eqs. (\ref{f in s
polar}) and (\ref{f* in s polar}), one finds after performing the integration with
respect to variables $k$ and $\varphi$
\begin{eqnarray} \label{Lz general}
L_z & = &
  \frac{2 \pi \varepsilon_0 l}{k^2 \omega} \delta(k-k') \int_0^{\pi}
  |f_l(\vartheta)|^2 \sin \vartheta d\vartheta
  +\frac{2 \pi \varepsilon_0 \sigma}{k^2 \omega} \delta(k-k') \nonumber \\ & &
  \times \int_0^{\pi} \left\{\frac{1}{2}\left(\frac{\cos \vartheta+\cos \Theta}{|\cos
  \vartheta+ \cos \Theta|}+ \frac{\cos \vartheta- \cos \Theta}{|\cos \vartheta- \cos
  \Theta|}\right)- \cos \vartheta \right\} |f_l(\vartheta)|^2 \sin \vartheta d\vartheta.
  \end{eqnarray}
In obtaining Eq. (\ref{Lz general}), the following integral formula is used:
\begin{equation} \label{equality}
\int_0^{\pi} \frac{dx}{1+a \cos x}= \frac{\pi}{\sqrt{1-a^2}}, \hspace{5pt} (|a|^2<1).
\end{equation}
Substituting Eq. (\ref{factorization}) into Eq. (\ref{W}) and considering Eqs. (\ref{f in
s polar}), (\ref{f* in s polar}), and (\ref{normalization}), one has for the total energy
of the beam
\begin{equation} \label{energy final}
W= \frac{2 \pi \varepsilon_0}{k^2} \delta(k-k') \int_0^{\pi} |f_l (\vartheta)|^2 \sin
\vartheta d\vartheta.
\end{equation}
It is clear that the orbital AM per unit energy is
\begin{equation} \label{Lz in W}
\frac{L_z}{W}= \frac{l}{\omega}+ \frac{\sigma}{\omega}
  \frac{\int_0^{\pi} \left\{\frac{1}{2}\left(\frac{\cos \vartheta+\cos \Theta}{|\cos
  \vartheta+ \cos \Theta|}+ \frac{\cos \vartheta- \cos \Theta}{|\cos \vartheta- \cos
  \Theta|}\right)- \cos \vartheta \right\} |f_l(\vartheta)|^2 \sin \vartheta d\vartheta}
  {\int_0^{\pi} |f_l(\vartheta)|^2 \sin \vartheta d\vartheta}.
\end{equation}

Next let us calculate the transverse component of orbital AM. The $x$-component is
rewritten from Eq. (\ref{L}) to be
\begin{equation} \label{Lx general}
L_x = -\int \frac{\varepsilon_0}{\omega} \mathbf{E}^{\dag}
  [k_y (i\frac{\partial}{\partial k_z})- k_z (i\frac{\partial}{\partial k_y})]
  \mathbf{E} k^2 \sin\vartheta dk d\vartheta d\varphi.
\end{equation}
According to Eq. (\ref{factorization}), one has
\begin{eqnarray} \label{Lx density}
\mathbf{E}^{\dag} [k_y (i\frac{\partial}{\partial k_z})- k_z (i\frac{\partial}{\partial
k_y})] \mathbf{E}
  & = &
  \tilde{\alpha}^{\dag} [k_y m^T (i\frac{\partial m}{\partial k_z})-
                         k_z m^T (i\frac{\partial m}{\partial k_y})] \tilde{\alpha} f^* f
  \nonumber \\
  & + & f^* [k_y (i\frac{\partial}{\partial k_z})- k_z (i\frac{\partial}{\partial k_y})] f.
\end{eqnarray}
When property (\ref{normalization}) is taken into account, straightforward calculations
yield
\begin{eqnarray} \label{differentiation}
k_y m^T (i \frac{\partial m}{\partial k_z})-k_z m^T (i \frac{\partial m}{\partial k_y}) &
= &
   \hat{\sigma}_3 \sin\vartheta \cos\varphi
  +\frac{\hat{\sigma}_3}{2} \frac{(\cos\vartheta-\cos\Theta)\cot\Theta}
   {1-\cos\Theta \cos\vartheta- \sin\Theta \sin\vartheta \cos\varphi} \nonumber \\
& &
  +\frac{\hat{\sigma}_3}{2} \frac{(\cos\vartheta+\cos\Theta) \cot\Theta}
   {1+\cos\Theta \cos\vartheta+ \sin\Theta \sin\vartheta \cos\varphi}.
\end{eqnarray}
Substituting Eqs. (\ref{Lx density}) and (\ref{differentiation}) into Eq. (\ref{Lx
general}) and considering the rotation symmetry of $f$ in Eq. (\ref{f in s polar}), one
obtains after performing the integration with respect to variables $k$ and $\varphi$,
\begin{equation} \label{Lx}
L_x=-\frac{\pi \varepsilon_0 \sigma}{k^2 \omega} \delta(k-k') \cot\Theta
   \int_0^{\pi} \left(\frac{\cos\vartheta+\cos\Theta}{|\cos\vartheta+\cos\Theta|}
                     +\frac{\cos\vartheta-\cos\Theta}{|\cos\vartheta-\cos\Theta|} \right)
  |f_l (\vartheta)|^2 \sin\vartheta d\vartheta.
\end{equation}
In obtaining Eq. (\ref{Lx}), formula (\ref{equality}) is used. The $x$-component of
orbital AM per unit energy is thus
\begin{equation} \label{Lx in W}
\frac{L_x}{W}=-\frac{\sigma \cot\Theta}{\omega}
   \frac{\int_0^{\pi} \frac{1}{2}
   \left(\frac{\cos\vartheta+\cos\Theta}{|\cos\vartheta+\cos\Theta|}
        +\frac{\cos\vartheta-\cos\Theta}{|\cos\vartheta-\cos\Theta|} \right)
  |f_l (\vartheta)|^2 \sin\vartheta d\vartheta}
  {\int_0^{\pi} |f_l(\vartheta)|^2 \sin \vartheta d\vartheta}.
\end{equation}
Similar calculations give for the $y$-component of orbital AM per unit energy
\begin{equation} \label{Ly in W}
\frac{L_y}{W}=0.
\end{equation}

Eqs. (\ref{Lz in W}), (\ref{Lx in W}), and (\ref{Ly in W}) are valid for any physically
allowed angular-spectrum function $f_l(\vartheta)$. Remembering that the unit vector
$\mathbf{I}$ lies in the plane $zox$, they show that as a vector quantity, the orbital AM
with respect to the origin is located in the plane formed by $\mathbf{I}$ and the
propagation direction for the rotation-symmetry electric scalar (\ref{f in s polar}).
Apart from an $l$-dependent term in the longitudinal component, the orbital AM is closely
dependent on the polarization $\sigma$ through the unit vector $\mathbf{I}$.

For a beam propagating in the $z$-direction, property (\ref{fl propagation}) is
satisfied. Considering our postulation (\ref{postulation}), Eqs. (\ref{Lz in W}) and
(\ref{Lx in W}) bocome
\begin{eqnarray} \label{Lz final}
\frac{L_z}{W}& = & \frac{l}{\omega}+ \frac{\sigma}{\omega}
  \frac{ \int_0^{\pi/2} \left\{\frac{1}{2}\left(1+ \frac{\cos \vartheta- \cos \Theta}{|\cos
  \vartheta- \cos \Theta|}\right)- \cos \vartheta \right\} |f_l(\vartheta)|^2 \sin \vartheta
  d\vartheta} {\int_0^{\pi/2} |f_l(\vartheta)|^2 \sin \vartheta d\vartheta}, \\
\label{Lx final}
\frac{L_x}{W}& = & -\frac{\sigma \cot\Theta}{\omega}
   \frac{\int_0^{\pi/2} \frac{1}{2}
   \left(1+\frac{\cos\vartheta-\cos\Theta}{|\cos\vartheta-\cos\Theta|} \right)
  |f_l (\vartheta)|^2 \sin\vartheta d\vartheta}
  {\int_0^{\pi/2} |f_l(\vartheta)|^2 \sin \vartheta d\vartheta},
\end{eqnarray}
respectively. Eq. (\ref{Lx final}) indicates that if $\mathbf{I}$ is neither
perpendicular nor parallel to the propagation direction, the transverse component of
orbital AM does not vanish. Let us discuss the following three cases.

\subsubsection{$|\Theta|=\frac{\pi}{2}$}

This is the case in which $\mathbf{I}$ is perpendicular to the propagation direction. The
beam described in this case is uniformly polarized \cite{Li1} in the paraxial
approximation in the traditional sense \cite{Pattanayak}. In this case, Eqs. (\ref{Lz
final}) and (\ref{Lx final}) become
\begin{eqnarray} \label{Lz pi/2}
\frac{L_z}{W} & = & \frac{l}{\omega}+ \frac{\sigma}{\omega}
  \frac{\int_0^{\pi/2} (1-\cos \vartheta) |f_l(\vartheta)|^2 \sin \vartheta
  d\vartheta} {\int_0^{\pi/2} |f_l(\vartheta)|^2 \sin \vartheta d\vartheta}, \\
\frac{L_x}{W}& = & 0 \nonumber,
\end{eqnarray}
respectively, indicating that the transverse component vanishes and the longitudinal
component depends on the polarization. It should be noted that the vanishing transverse
component here is just with respect to the origin. With respect to any reference point
that is not on the beam axis (the $z$-axis), the transverse component is by no means
equal to zero as is shown by Eq. (\ref{OAM wrt x0}). Furthermore, by making use of
paraxial approximation in which $\cos \vartheta$ in the integrand of the numerator can be
approximated by unity, $\cos \vartheta \approx 1$, Eq. (\ref{Lz pi/2}) reduces to
\begin{equation} \label{Lz pi/2 paraxial}
\frac{L_z}{W}= \frac{l}{\omega}.
\end{equation}
Only under so special conditions, is the longitudinal component of orbital AM
approximately independent of the polarization. Eq. (\ref{Lz pi/2 paraxial}) is exactly
the result that was obtained from the consideration of paraxial Laguerre-Gaussian beams
\cite{Allen-BSW}.

\subsubsection{$\Theta=0$}

This is the case in which the unit vector $\mathbf{I}$ is parallel to the propagation
direction. The beam described in this case is known as cylindrical vector beam
\cite{Youngworth, Li2}. In this case, Eqs. (\ref{Lz final}) and (\ref{Lx final}) become
\begin{eqnarray} \label{Lz zero}
\frac{L_z}{W}& = & \frac{l}{\omega}- \frac{\sigma}{\omega}
  \frac{ \int_0^{\pi/2} |f_l(\vartheta)|^2 \cos \vartheta \sin \vartheta
  d\vartheta} {\int_0^{\pi/2} |f_l(\vartheta)|^2 \sin \vartheta d\vartheta}, \\
\frac{L_x}{W}& = & 0 \nonumber,
\end{eqnarray}
respectively. The transverse component vanishes too. But it is seen from Eq. (\ref{Lz
zero}) that even in the paraxial approximation, the longitudinal component is not
independent of the polarization and is given by
\begin{equation} \label{Lz zero paraxial}
\frac{L_z}{W}= \frac{l}{\omega}- \frac{\sigma}{\omega}.
\end{equation}

\subsubsection{$|\Theta| \gg \Delta \vartheta$}

A well-collimated beam has a very narrow divergence angle $\Delta \vartheta$. This
situation allows us to consider such a case in which $|\Theta| \gg \Delta \vartheta$ is
satisfied. The refracted beam that occurred in the spin Hall effect of light
\cite{Hosten-K} was proven \cite{Li3} to belong to this category. In this case, we have
$\cos\vartheta- \cos\Theta>0$ in the region in which $|f_l (\vartheta)|$ is appreciable.
Eqs. (\ref{Lz final}) and (\ref{Lx final}) are thus approximated as
\begin{eqnarray}
\frac{L_z}{W} & \approx & \frac{l}{\omega}+ \frac{\sigma}{\omega}
  \frac{ \int_0^{\pi/2} (1-\cos \vartheta) |f_l(\vartheta)|^2 \sin \vartheta
  d\vartheta} {\int_0^{\pi/2} |f_l(\vartheta)|^2 \sin \vartheta d\vartheta}, \nonumber\\
\label{Lx large Theta}
\frac{L_x}{W}& \approx & -\frac{\sigma \cot\Theta}{\omega},
\end{eqnarray}
respectively. The longitudinal component is almost equal to that in the case of
$|\Theta|=\frac{\pi}{2}$. But the transverse component is not equal to zero. Eq. (\ref{Lx
large Theta}) expresses a simple polarization dependence through the unit vector
$\mathbf{I}$.

\subsection{Spin is dependent only on the polarization}

Substituting Eq. (\ref{factorization}) into Eq. (\ref{S}) and taking Eqs. (\ref{f in s
polar}) and (\ref{f* in s polar}) into account, one gets
\begin{equation}
\mathbf{S}= \frac{\varepsilon_0 \sigma}{k^2 \omega} \delta(k-k') \int
\frac{\mathbf{k}}{k} |f_l (\vartheta)|^2 \sin \vartheta d\vartheta d\varphi.
\end{equation}
It shows that the transverse component of spin vanishes. The longitudinal component is
given by
\begin{equation} \label{Sz}
S_z= \frac{2 \pi \varepsilon_0 \sigma}{k^2 \omega} \delta(k-k') \int_0^{\pi} |f_l
(\vartheta)|^2 \cos\vartheta \sin \vartheta d\vartheta.
\end{equation}
Clearly, the spin AM does not depend on the unit vector $\mathbf{I}$. From Eqs.
(\ref{Sz}) and (\ref{energy final}), it follows that the longitudinal component of spin
per unit energy is
\begin{equation} \label{Sz in W}
\frac{S_z}{W}= \frac{\sigma}{\omega} \frac{\int_0^{\pi} |f_l (\vartheta)|^2 \cos\vartheta
\sin \vartheta d\vartheta}{\int_0^{\pi} |f_l (\vartheta)|^2 \sin \vartheta d\vartheta},
\end{equation}
which is valid for any physically allowed angular-spectrum function $f_l(\vartheta)$. For
a paraxial beam, $\cos \vartheta \approx 1$ holds and Eq. (\ref{Sz in W}) reduces to
\begin{equation} \label{Sz paraxial}
\frac{S_z}{W} \approx \frac{\sigma}{\omega}.
\end{equation}
This is what was obtained from the consideration of paraxial Laguerre-Gaussian beams
\cite{Allen-BSW}.

\subsection{Total AM}

The total AM is the sum of spin and orbital AM. Since the transverse component of spin
vanishes, we discuss here only the property of longitudinal component of the total AM.
Combining Eqs. (\ref{Lz in W}) and (\ref{Sz in W}) together, one has
\begin{equation} \label{Jz in W}
\frac{J_z}{W}= \frac{l}{\omega}+ \frac{\sigma}{\omega}
  \frac{\int_0^{\pi} \frac{1}{2}\left(\frac{\cos \vartheta+\cos \Theta}{|\cos \vartheta+
  \cos \Theta|}+ \frac{\cos \vartheta- \cos \Theta}{|\cos \vartheta- \cos \Theta|}\right)
  |f_l(\vartheta)|^2 \sin \vartheta d\vartheta}
  {\int_0^{\pi} |f_l(\vartheta)|^2 \sin \vartheta d\vartheta}.
  \end{equation}
It is instructive to note that $J_z$ does consist of two parts. One depends only on an
integer $l$, and the other depends only on $\sigma$. But the former is not the orbital
AM, and the latter is not the spin AM. Eq. (\ref{Jz in W}) is valid for any physically
allowed function $f_l(\vartheta)$. When Eq. (\ref{fl propagation}) is taken into account
for a beam propagating in the $z$-direction, it becomes
\begin{equation} \label{Jz final}
\frac{J_z}{W}= \frac{l}{\omega}+ \frac{\sigma}{\omega}
  \frac{\int_0^{\pi/2} \frac{1}{2}\left(1+ \frac{\cos \vartheta- \cos \Theta}{|\cos
  \vartheta- \cos \Theta|}\right) |f_l(\vartheta)|^2 \sin \vartheta d\vartheta}
  {\int_0^{\pi/2} |f_l(\vartheta)|^2 \sin \vartheta d\vartheta},
\end{equation}
which clearly shows the impact of the unit vector $\mathbf{I}$. If $\Theta=0$, Eq.
(\ref{Jz final}) reduces to
\begin{equation} \label{Jz zero}
\frac{J_z}{W}= \frac{l}{\omega},
\end{equation}
which is independent of the polarization whether the beam is paraxial or not. If
$|\Theta|= \frac{\pi}{2}$ on the other hand, one gets from Eq. (\ref{Jz final})
\begin{equation} \label{Jz pi/2}
\frac{J_z}{W}= \frac{l}{\omega}+ \frac{\sigma}{\omega},
\end{equation}
which is also valid beyond the paraxial approximation. Though the total AM exhibits so
simple dependence on $l$ and $\sigma$, the first term $\frac{l}{\omega}$ is not the
orbital AM and the second one $\frac{\sigma}{\omega}$ is not the spin AM, unless the
paraxial approximation holds. It will be shown in the next section that the polarization
dependence of $L_z$ for a nonparaxial beam of perpendicular $\mathbf{I}$ is the basis of
conversion from spin to orbital AM by a high numerical aperture.

In summary of this section, I have shown that the orbital AM is closely related to the
unit vector $\mathbf{I}$. It is due to the impact of $\mathbf{I}$ that the orbital AM is
dependent on the polarization. If $\mathbf{I}$ is parallel to the propagation direction,
both the spin and orbital AM have only longitudinal components. They are all polarization
dependent whether the beam is paraxial or not. But the total AM does not depend on the
polarization. To the best of my knowledge, this is the first time to give the AM
expression of cylindrical vector beams. If $\mathbf{I}$ is perpendicular to the
propagation direction, the spin and orbital AM also have only longitudinal components.
But in the paraxial approximation, the orbital AM is nearly independent of the
polarization and is equal to $\frac{l}{\omega}$, and the spin AM is nearly equal to
$\frac{\sigma}{\omega}$. If $\mathbf{I}$ is neither parallel nor perpendicular to the
propagation direction, the transverse component of orbital AM is not equal to zero.
Comparison with the result of Ref. \cite{Allen-BSW} indicates that the unit vector
$\mathbf{I}$ of Laguerre-Gaussian beams is perpendicular to the propagation direction.

\section{\label{Applications} Applications}

In this section, I will apply the results obtained before to discuss two different
problems. One is the so-called paradox on the spin of circularly polarized plane wave. It
will be shown that such a paradox does not exist at all. The other is the conversion of
partial spin of a paraxial beam to the orbital AM of the focused beam through a high
numerical aperture. The conversion will be shown to root in the polarization dependence
of both spin and orbital AM.

\subsection{There is no paradox on the spin of circularly polarized plane wave}

The so-called paradox on the spin of circularly polarized plane wave has been the subject
of discussion \cite{Humblet, Jauch, Simmons} ever since Beth \cite{Beth} experimentally
demonstrated that a circularly polarized plane wave carries spin AM $\hbar$ and was still
investigated recently \cite{Allen-P, Khrapko, Allen-P1, Stewart, Masud}. It states that
because the electric and magnetic vectors of a circularly polarized plane wave are
perpendicular to the wave vector, its momentum density must be in the propagation
direction. As a result, the AM component in the propagation direction must be zero
\cite{Lenstra} due to the cross product of the position vector with the momentum density.
This is contrary to Beth's observation.

As we have shown in Section \ref{AM separation}, the spin of an electromagnetic field in
free space does not come from the part of momentum density that produces the net
momentum. Instead, it originates from the other part of momentum density that does not
have contribution to the net momentum. From this point of view, it follows that there is
no paradox on the spin of circularly polarized plane wave. After all, what is produced
from the momentum density in the propagation direction is the net momentum. In order to
elucidate that the spin does not originate from this momentum density, let us make use of
Eq. (\ref{S}) to calculate the AM of a plane wave.

The electric vector of a plane wave in reciprocal space is given by
\begin{equation} \label{pw spectrum}
\mathbf{E}= m \tilde{\alpha} f_0 \delta^3 (\mathbf{k}-\mathbf{k}'),
\end{equation}
where $\mathbf{k}'$ is the wave vector of the plane wave. If Eq. (\ref{pw spectrum}) is
substituted directly into Eq. (\ref{S}), an infinity will occur. To deal with the
infinity, we make use of the technique of $\delta$-normalization as before by replacing
$\mathbf{E}^*$ with
\begin{equation} \label{conjugate pw spectrum}
\mathbf{E}^*= m \tilde{\alpha}^* f^*_0 \delta^3 (\mathbf{k}-\mathbf{k}'').
\end{equation}
Substituting Eqs. (\ref{pw spectrum}) and (\ref{conjugate pw spectrum}) into Eq.
(\ref{S}), one gets
\begin{equation} \label{S plane wave}
\mathbf{S}=\frac{\sigma}{\omega} \varepsilon_0 |f_0|^2 \frac{\mathbf{k}}{k} \delta^3
(\mathbf{k}-\mathbf{k}').
\end{equation}
Similarly, substituting Eqs. (\ref{pw spectrum}) and (\ref{conjugate pw spectrum}) into
Eq. (\ref{W}), one has for the total energy of the wave
\begin{equation} \label{W plane wave}
W= \varepsilon_0 |f_0|^2 \delta^3 (\mathbf{k}-\mathbf{k}').
\end{equation}
It follows that the spin per photon in the plane wave is
\begin{equation} \label{spin per photon}
\frac{\mathbf{S}}{W} \hbar \omega= \hbar \sigma \frac{\mathbf{k}}{k},
\end{equation}
which is entirely along the direction of wave vector $\mathbf{k}$. For circular
polarizations $\sigma= \pm 1$, the spin AM per photon is $\pm \hbar$, which is in perfect
agreement with Beth's experimental observation. This indicates that when one talked about
the paradox on the plane wave's spin, he/she did not realize the role that the momentum
density in Eq. (\ref{part 2 of p}) plays in the AM. It is very interesting to note that
we arrive at the quantum feature \cite{Mandel} of photon's spin by a purely classical
approach, from which one might appreciate the nonlocal property of the photon. Since the
spin comes from the momentum density that does not produce any momentum on one hand and
is stored in the whole real space over which the plane wave spreads on the other, it
might be probable that the concept of photon's spin density in real space is physically
meaningless \cite{Allen-P}.

\subsection{Conversion from spin to orbital AM by a high numerical aperture}

The incident beam in the AM conversion experiment \cite{Zhao} is LG$_0^1$, a
Laguerre-Gaussian beam. So its unit vector $\mathbf{I}$ is perpendicular to the
propagation direction and its parameter $l$ is equal to one, $l=1$. Before focusing, the
spin and orbital AM per unit energy of the paraxial beam in the propagation direction are
approximately $\frac{\sigma} {\omega}$ and $\frac{1}{\omega}$, respectively, as Eqs.
(\ref{Sz paraxial}) and (\ref{Lz pi/2 paraxial}) show. After focusing, the spin per unit
energy of the non-paraxial beam is obtained from Eq. (\ref{Sz in W}) to be
$$
\frac{\sigma}{\omega} \frac{\int_0^{\pi/2} |f_l (\vartheta)|^2 \cos\vartheta \sin
\vartheta d\vartheta}{\int_0^{\pi/2} |f_l (\vartheta)|^2 \sin \vartheta d\vartheta},
$$
indicating that only a fraction of the incident spin remains in the focused beam, where
$f_l (\vartheta)$ now stands for the angular-spectrum function of the focused beam. If
the rest of the incident spin
$$
\frac{\sigma}{\omega} \left(1- \frac{\int_0^{\pi/2} |f_l(\vartheta)|^2 \cos\vartheta
   \sin\vartheta d\vartheta} {\int_0^{\pi/2} |f_l(\vartheta)|^2 \sin\vartheta
   d\vartheta}\right)
$$
is converted into the orbital AM \cite{Nieminen}, the orbital AM of the focused beam
should be
$$
\frac{1}{\omega}+ \frac{\sigma}{\omega}
   \left(1- \frac{\int_0^{\pi/2} |f_l(\vartheta)|^2 \cos\vartheta \sin\vartheta
  d\vartheta} {\int_0^{\pi/2} |f_l(\vartheta)|^2 \sin\vartheta d\vartheta}\right).
$$
This is just the result predicted by Eq. (\ref{Lz pi/2}). We thus explain the conversion
from the spin to the orbital AM on the basis that the orbital AM can be dependent on the
polarization. If $\sigma=-1$, the orbital AM per photon is less than $\hbar$. On the
other hand, if $\sigma=1$, the orbital AM per photon is larger than $\hbar$. The authors
of Ref. \cite{Zhao} put forward their own theoretical explanation based on the analysis
of the longitudinal component of the focused beam's electric vector. Because the
longitudinal component of the electric vector is not able to represent the whole beam,
they failed to show how the orbital AM of the focused beam depends on the polarization of
the incident paraxial beam.

\section{Conclusions and remarks} \label{conclusions}

In conclusion, I put forward a rigorous approach to the separation of the total AM into
the spin and orbital AM. This approach is based on the analysis of the momentum density.
It was shown that the momentum density can split into two parts. One part that does not
produce any momentum corresponds to the spin. The other part that produces all the
momentum corresponds to the orbital AM. The spin defined this way was applied to show
that there is no paradox about the spin of circularly polarized plane wave. Apart from
the conclusion that the spin is dependent on the polarization, I further showed that the
orbital AM is also dependent on the polarization. The polarization-dependent orbital AM
was applied to explain the experiment \cite{Zhao} that converted partial spin of the
paraxial beam LG$_0^1$ into the orbital AM of the focused beam through a high numerical
aperture.

The unit vector $\mathbf{I}$ was shown to have evident impact on the orbital AM. In the
first place, Eqs. (\ref{Lz in W}), (\ref{Lx in W}), and (\ref{Ly in W}) show that the
orbital AM is located in the plane formed by $\mathbf{I}$ and the propagation direction.
Secondly, Eqs. (\ref{derivative}) and (\ref{differentiation}) show that the
polarization-dependent term of orbital AM is determined by the direction of $\mathbf{I}$.
when $\mathbf{I}$ is parallel to the propagation direction, the orbital AM is always
dependent on the polarization. When $\mathbf{I}$ is perpendicular to the propagation
direction, the orbital AM is almost independent of the polarization in the paraxial
approximation. These phenomena may imply that the orbital AM is most connected with the
polarization, the inner degree of freedom, when $\mathbf{I}$ is parallel to the
propagation direction and is least connected with the inner degree of freedom when
$\mathbf{I}$ is perpendicular to the propagation direction. In a word, the impact of
$\mathbf{I}$ on the orbital AM may offer further insights into the nature of the AM of
light.

The author would like to thank Masud Mansuripur of the University of Arizona and Thomas
G. Brown of the University of Rochester for their helpful discussions. This work was
supported in part by the National Natural Science Foundation of China (60877055 and
60806041), the Science and Technology Commission of Shanghai Municipal (08JC1409701 and
08QA14030), the Shanghai Educational Development Foundation (2007CG52), and the Shanghai
Leading Academic Discipline Project (S30105).

\appendix*

\section{Derivation of Eqs. (\ref{S}) and (\ref{L})}

Let us first derive Eq. (\ref{L}). Substituting Eq. (\ref{part 1 of p}) into Eq.
(\ref{total OAM}), one has
\begin{equation} \label{L parts}
\mathbf{L}(0)= \mathbf{L}_1+ \mathbf{L}_2+ c.c.,
\end{equation}
where
\begin{equation} \label{L1}
\mathbf{L}_1= \frac{\varepsilon_0}{4(2 \pi)^3} \int d^3 k' d^3 k \int d^3 x \frac{\mathbf{E}'
\cdot \mathbf{E}}{\omega} \mathbf{x} \times \mathbf{k} e^{i(\mathbf{k}'+ \mathbf{k}) \cdot
\mathbf{x}} e^{-i(\omega'+ \omega)t},
\end{equation}
and
\begin{equation} \label{L2}
\mathbf{L}_2= \frac{\varepsilon_0}{4(2 \pi)^3} \int d^3 k' d^3 k \int d^3 x \frac{\mathbf{E}'
\cdot \mathbf{E}^*}{\omega} \mathbf{x} \times \mathbf{k} e^{i(\mathbf{k}'- \mathbf{k}) \cdot
\mathbf{x}} e^{-i(\omega'- \omega)t}.
\end{equation}
Upon integrating Eq. (\ref{L1}) over the real space and noticing the following properties
of Dirac's $\delta$ function and its first-order derivative,
\begin{equation} \label{delta 1}
\delta(t)=\frac{1}{2 \pi} \int_{-\infty}^{\infty} \exp(i \omega t) d \omega, \hspace{5pt}
\delta'(t)=\frac{i}{2 \pi} \int_{-\infty}^{\infty} \omega \exp(i \omega t) d \omega,
\end{equation}
one obtains
\begin{eqnarray}
\mathbf{L}_1 & = & \frac{\varepsilon_0}{4i} \int (k_y \mathbf{e}_z-k_z \mathbf{e}_y)
          \frac{\mathbf{E}' \cdot \mathbf{E}}{\omega} e^{-i (\omega'+ \omega)t} \delta' (k'_x+ k_x)
          \delta (k'_y+ k_y) \delta (k'_z+ k_z) d^3 k' d^3 k \nonumber \\
    & + & \frac{\varepsilon_0}{4i} \int (k_z \mathbf{e}_x-k_x \mathbf{e}_z)
          \frac{\mathbf{E}' \cdot \mathbf{E}}{\omega} e^{-i (\omega'+ \omega)t} \delta (k'_x+ k_x)
          \delta' (k'_y+ k_y) \delta (k'_z+ k_z) d^3 k' d^3 k \nonumber \\
    & + & \frac{\varepsilon_0}{4i} \int (k_x \mathbf{e}_y-k_y \mathbf{e}_x)
          \frac{\mathbf{E}' \cdot \mathbf{E}}{\omega} e^{-i (\omega'+ \omega)t} \delta (k'_x+ k_x)
          \delta (k'_y+ k_y) \delta' (k'_z+ k_z) d^3 k' d^3 k. \nonumber
\end{eqnarray}
It is changed by eliminating the $\delta$ functions into
\begin{eqnarray}
\mathbf{L}_1 & = & \frac{\varepsilon_0}{4i} \int (k_y \mathbf{e}_z-k_z \mathbf{e}_y)
          \frac{\mathbf{E}(k'_x, -k_y, -k_z) \cdot \mathbf{E}}{\omega} e^{-i [(\omega(k'_x, -k_y, -k_z)+
          \omega]t} \delta' (k'_x+ k_x) dk'_x d^3 k \nonumber \\
    & + & \frac{\varepsilon_0}{4i} \int (k_z \mathbf{e}_x-k_x \mathbf{e}_z)
          \frac{\mathbf{E}(-k_x, k'_y, -k_z) \cdot \mathbf{E}}{\omega} e^{-i [\omega(-k_x, k'_y, -k_z)+
          \omega]t} \delta' (k'_y+ k_y) d k'_y d^3 k \nonumber \\
    & + & \frac{\varepsilon_0}{4i} \int (k_x \mathbf{e}_y-k_y \mathbf{e}_x)
          \frac{\mathbf{E}(-k_x, -k_y, k'_z) \cdot \mathbf{E}}{\omega} e^{-i [\omega(-k_x, -k_y, k'_z)+
          \omega]t} \delta' (k'_z+ k_z) d k'_z d^3 k. \nonumber
\end{eqnarray}
Noticing the following property of the derivative of the $\delta$ function,
\begin{equation} \label{delta 2}
\int_{t_1}^{t_2} f(t) \delta' (t-t_0) dt= -f'(t_0), \hspace{5pt} t_1<t_0<t_2,
\end{equation}
and taking transformation (\ref{Hermitian}) into account, the above equation is reduced to
\begin{eqnarray}
\mathbf{L}_1 & = & \frac{\varepsilon_0}{4i} \int \frac{k_y \mathbf{e}_z-k_z \mathbf{e}_y}{\omega}
          \left(\mathbf{E} \cdot \frac{\partial \mathbf{E}^*}{\partial k_x}+
          i \frac{k_x t}{\varepsilon_0 \mu_0 \omega} \mathbf{E}^* \cdot \mathbf{E} \right) d^3 k
          \nonumber \\
    & + & \frac{\varepsilon_0}{4i} \int \frac{k_z \mathbf{e}_x-k_x \mathbf{e}_z}{\omega}
          \left(\mathbf{E} \cdot \frac{\partial \mathbf{E}^*}{\partial k_y}+
          i \frac{k_y t}{\varepsilon_0 \mu_0 \omega} \mathbf{E}^* \cdot \mathbf{E} \right) d^3 k
          \nonumber \\
    & + & \frac{\varepsilon_0}{4i} \int \frac{k_x \mathbf{e}_y-k_y \mathbf{e}_x}{\omega}
          \left(\mathbf{E} \cdot \frac{\partial \mathbf{E}^*}{\partial k_z}+
          i \frac{k_z t}{\varepsilon_0 \mu_0 \omega} \mathbf{E}^* \cdot \mathbf{E} \right) d^3 k
          \nonumber \\
    & = & \frac{i \varepsilon_0}{4} \int \frac{1}{\omega} \mathbf{E}^T (\mathbf{k} \times
          \nabla_{\mathbf{k}}) \mathbf{E}^* d^3 k. \nonumber
\end{eqnarray}
By making the variable replacement $\mathbf{k} \rightarrow -\mathbf{k}$, it is changed
into a familiar form,
\begin{equation} \label{L1 ex}
\mathbf{L}_1= \frac{1}{4} \int \frac{\varepsilon_0}{i \omega} \mathbf{E}^{\dag}
(\mathbf{k} \times \nabla_{\mathbf{k}}) \mathbf{E} d^3 k.
\end{equation}
Since operator $-i \nabla_{\mathbf{k}}$ is Hermitian, the $\mathbf{L}_1$ in Eq. (\ref{L1
ex}) is real. A similar calculation produces from Eq. (\ref{L2})
\begin{equation} \label{L2 ex}
\mathbf{L}_2= \mathbf{L}_1.
\end{equation}
It is clear that substituting Eqs. (\ref{L1 ex}) and (\ref{L2 ex}) into Eq. (\ref{L parts}) will
yield Eq. (\ref{L}).

Then we derive Eq. (\ref{S}). Substituting Eq. (\ref{part 2 of p}) into Eq. (\ref{total
SAM}), one has
\begin{equation} \label{S parts}
\mathbf{S}= \mathbf{S}_1+ \mathbf{S}_2+ c.c.,
\end{equation}
where
\begin{equation} \label{S1}
\mathbf{S}_1= -\frac{\varepsilon_0}{4(2 \pi)^3} \int d^3 k' d^3 k \int d^3 x \frac{\mathbf{E}'
\cdot \mathbf{k}}{\omega} \mathbf{x} \times \mathbf{E} e^{i(\mathbf{k}'+ \mathbf{k}) \cdot
\mathbf{x}} e^{-i(\omega'+ \omega)t},
\end{equation}
and
\begin{equation} \label{S2}
\mathbf{S}_2= -\frac{\varepsilon_0}{4(2 \pi)^3} \int d^3 k' d^3 k \int d^3 x \frac{\mathbf{E}'
\cdot \mathbf{k}}{\omega} \mathbf{x} \times \mathbf{E}^* e^{i(\mathbf{k}'- \mathbf{k}) \cdot
\mathbf{x}} e^{-i(\omega'- \omega)t}.
\end{equation}
Upon integrating Eq. (\ref{S1}) over the real space and noticing Eq. (\ref{delta 1}), one
obtains
\begin{eqnarray}
\mathbf{S}_1
    & = & \frac{i \varepsilon_0}{4} \int (E_y \mathbf{e}_z-E_z \mathbf{e}_y)
          \frac{\mathbf{E}' \cdot \mathbf{k}}{\omega} e^{-i (\omega'+ \omega)t} \delta' (k'_x+ k_x)
          \delta (k'_y+ k_y) \delta (k'_z+ k_z) d^3 k' d^3 k \nonumber \\
    & + & \frac{i \varepsilon_0}{4} \int (E_z \mathbf{e}_x-E_x \mathbf{e}_z)
          \frac{\mathbf{E}' \cdot \mathbf{k}}{\omega} e^{-i (\omega'+ \omega)t} \delta (k'_x+ k_x)
          \delta' (k'_y+ k_y) \delta (k'_z+ k_z) d^3 k' d^3 k \nonumber \\
    & + & \frac{i \varepsilon_0}{4} \int (E_x \mathbf{e}_y-E_y \mathbf{e}_x)
          \frac{\mathbf{E}' \cdot \mathbf{k}}{\omega} e^{-i (\omega'+ \omega)t} \delta (k'_x+ k_x)
          \delta (k'_y+ k_y) \delta' (k'_z+ k_z) d^3 k' d^3 k. \nonumber
\end{eqnarray}
It is changed into, by eliminating the $\delta$ functions and taking Eqs. (\ref{delta 2}) and
(\ref{Hermitian}) into account,
\begin{eqnarray}
\mathbf{S}_1 & = &
  \frac{i \varepsilon_0}{4} \int \frac{E_y \mathbf{e}_z-E_z
    \mathbf{e}_y}{\omega} \mathbf{k} \cdot \frac{\partial \mathbf{E}^*}{\partial k_x} d^3 k
 +\frac{i \varepsilon_0}{4} \int \frac{E_z \mathbf{e}_x-E_x \mathbf{e}_z}{\omega}
    \mathbf{k} \cdot \frac{\partial \mathbf{E}^*}{\partial k_y} d^3 k \nonumber \\
 & + & \frac{i \varepsilon_0}{4} \int \frac{E_x \mathbf{e}_y-E_y \mathbf{e}_x}{\omega}
    \mathbf{k} \cdot \frac{\partial \mathbf{E}^*}{\partial k_z} d^3 k. \nonumber
\end{eqnarray}
From the transversality condition $\mathbf{k} \cdot \mathbf{E}^*=0$, we know that
$$
  \mathbf{k} \cdot \frac{\partial \mathbf{E}^*}{\partial k_x}=-E_x^*, \hspace{5pt}
  \mathbf{k} \cdot \frac{\partial \mathbf{E}^*}{\partial k_y}=-E_y^*, \hspace{5pt}
  \mathbf{k} \cdot \frac{\partial \mathbf{E}^*}{\partial k_z}=-E_z^*. \hspace{5pt}
$$
$\mathbf{S}_1$ then reduces to
\begin{equation} \label{S1 ex}
\mathbf{S}_1= \frac{1}{4} \int \frac{\varepsilon_0}{i\omega} \mathbf{E}^* \times \mathbf{E} d^3 k,
\end{equation}
which is clearly real. Similarly, $\mathbf{S}_2$ in Eq. (\ref{S2}) is found to be real and is
equal to $\mathbf{S}_1$,
\begin{equation} \label{S2 ex}
\mathbf{S}_2=\mathbf{S}_1.
\end{equation}
Substituting Eqs. (\ref{S1 ex}) and (\ref{S2 ex}) into Eq. (\ref{S parts}) will yield Eq.
(\ref{S}).

\end{document}